\documentclass[12pt]{article}
\usepackage[latin1]{inputenc}
\usepackage[dvips]{graphicx}
\usepackage{graphicx}
\newcommand{\ug}{\; = \;}

\newcommand{\pa}{\partial}

\newcommand{\text}{\rm}

\newcommand{\drm}{{\rm d}}

\newcommand{\infi}{\infty}

\newcommand{\kr}{k_{\rho}}

\newcommand{\bb}{\begin{equation}}
\newcommand{\ee}{\end{equation}}
\newcommand{\bega}{\begin{eqarray}}
\newcommand{\ega}{\end{eqnarray}}
\newcommand{\begae}{\begin{eqnarray*}}
\newcommand{\egae}{\end{eqnarray*}}

\newcommand{\h}{\hspace*{4ex}}
\newcommand{\dis}{\displaystyle}

\newcommand{\om}{\omega}

\newcommand{\cent}{\centerline}
\newcommand{\vs}{\vspace*}

\begin{document}

\baselineskip 0.8cm

\begin{center}

{\Large {\bf Unidirectional decomposition method for obtaining
exact localized waves solutions totally free of backward
components.}}

\end{center}

\vs{3mm}

\cent{ Michel Zamboni-Rached, }

\vs{0.1 cm}

\centerline{{\em Centro de Ciências Naturais e Humanas,
Universidade Federal do ABC, Santo Andr\'e, SP, Brasil.}}

\vs{0.5 cm}

\

\begin{abstract}

In this paper we use a unidirectional decomposition capable of
furnishing localized wave pulses, with luminal and superluminal
peak velocities, in exact form and totally free of backward
components, which have been a chronic problem for such wave
solutions. This decomposition is powerful enough for yielding not
only ideal nondiffracting pulses but also their finite energy
versions still in exact analytical closed form. Another advantage
of the present approach is that, since the backward spectral
components are absent, the frequency spectra of the pulses do not
need to possess ultra-widebands, as it is required by the usual
localized waves (LWs) solutions obtained by other methods.
Finally, the present results bring the LW theory nearer to the
real experimental possibilities of usual laboratories.

\end{abstract}

.\hfill\break
%%{\em OCIS codes\/}: \ 320.5550 ; \ 320.5540 .\\


\section{Introduction}

\h Localized waves (or nondiffracting waves)[1-60] are very
special free space solutions of the linear wave equation,
$(\nabla^2 -\partial_t^2)\psi = 0 $, whose main characteristic is
that of resisting the diffraction effects for long distances. In
their pulse versions, the LWs can possess subluminal, luminal or
supeluminal peak velocities. As it is well
known\cite{BSZ,SBZ,Donnelly,PIER98,MRH}, these impressive features
of the nondiffracting waves are due to the special space-time
coupling of their spectra.

\h In the last years, several methods\cite{BSZ,Tesi,PIER98,MRH}
have been developed to yield localized pulses in exact analytical
closed form. The most successful approaches are those dealing with
bidirectional or unidirectional
decomposition\cite{BSZ,PIER98,MRH}, in which the time variable $t$
and the spatial variable $z$ (considered as the propagation
direction) are replaced with other two variables, linear
combinations of them.

\h Examples of bidirectional and unidirectional decompositions are
i) $\zeta = z-ct$, $\eta=z+ct$, developed by Besieris
et.al.\cite{BSZ}; ii) $\zeta = z-Vt$, $\eta=z-c^2t/V$, also
introduced by those authors\cite{PIER98}; iii) $zeta = z-Vt$,
$\eta=z+Vt$, developed by Zamboni-Rached et.al.\cite{MRH}; iv)
$\zeta = z-Vt$, $\eta=z+ct$, proposed by Besieris
et.al.\cite{PIER98}.

\h Even if all those decompositions allow us to get exact
analytical LWs solutions, all of them suffer with the same
problem, that is, the occurrence of backward travelling components
in their spectra\footnote{Only the simplest X-wave pulses, which
possess the form $\psi(\rho,\phi,z-Vt)$, does not present this
problem.}. This drawback makes it necessary the use of
ultra-wideband frequency spectra to minimize\cite{BSZ,PIER98,MRH}
the contribution of the backward components. This fact can suggest
the wrong idea that ultra-wide frequency bands are a
characteristic of the LW pulses, which is not true. As a matter of
fact, it is quite simple to choose spectra that eliminate
completely such ''noncausal" components and at the same time have
narrow frequency bands. The real problem is that no closed form
analytical solution is known for those cases, and one has to make
recourse to time consuming numerical simulations. This problem was
already solved in the case of localized subluminal waves\cite{pra}
but still persists in the general cases of luminal and
superluminal LW pulses.

\h We are going to show here that, by a very simple decomposition,
a unidirectional one, one can overcome the problems cited above,
getting LW pulses with superluminal and luminal peak velocities,
constituted by forward\footnote{In the following, we shall briefly
write ''forward components" and even ''forward pulses".}
travelling components only, and without mandatory recourses to
ultra wide frequency bands. These results make nondiffracting
waves more easily experimentally realizable and applicable.

\section{The method}

\h The first subsection forwards a brief overview of the LWs
theory. The rest of this section is devoted to the new method and
to its results.

\subsection{A brief overview about LWs}

\h In the case of the linear and homogeneous wave equation in free
space, in cylindrical coordinates $(\rho,\phi,z)$ and using a
Fourier-Bessel expansion, we can express a general solution
$\psi(\rho,\phi,z,t)$ as

\bb \Psi(\rho,\phi,z,t) \ug
\sum_{n=-\infty}^{\infty}\left[\int_{0}^{\infty}d\kr\,\int_{-\infi}^{\infty}dk_z\,\int_{-\infty}^{\infty}d\om\,\kr
A_n^{'}(\kr,k_z,\om) J_n(\kr \rho)e^{ik_z z}e^{-i\om t}e^{i n
\phi} \right] \label{S2geral1} \ee

with

\bb A_n^{'}(\kr,k_z,\om) \ug A_n(k_z,\om)\,\delta\left(\kr^2 -
\left(\frac{\om^2}{c^2} - k_z^2 \right)\right)
\label{S2sepc1}\,\,\, , \ee

$A_n(k_z,\om)$ being an arbitrary function and $\delta(.)$ the
Dirac's delta function.

\h An ideal nondiffracting wave can be defined as a wave capable
of maintaining its spatial form indefinitely (except for local
variations) while propagating. This property can be expressed in a
mathematical way\cite{Tesi} by (when assuming propagation in the
$z$ direction)

\bb \Psi(\rho,\phi,z,t) \ug \Psi(\rho,\phi,z + \Delta z_0,t +
\frac{\Delta z_0}{\mathcal{V}}) \label{S2def} \ee

where $\Delta z_0$ is a certain length and $\mathcal{V}$ is the
pulse-peak velocity, with $0\leq \mathcal{V} \leq \infty$.

\h Using (\ref{S2def}) in (\ref{S2geral1}), and taking into
account (\ref{S2sepc1}), we can show\cite{Tesi,Livro} that any
localized wave solution, when eliminating evanescent waves and
considering only positive angular frequencies, can be written as

\bb \begin{array}{clcr} \Psi(\rho,\phi,z,t) \ug &
\dis{\sum_{n=-\infty}^{\infty}\,\sum_{m=-\infty}^{\infty}}\left[\dis{\int_{0}^{\infty}\drm\om\,\int_{-\om/c}^{\om/c}\drm
k_z
A_{nm}(k_z,\om)}\right. \\

\\

& \times \left.J_n\left(\rho\sqrt{\dis{\frac{\om^2}{c^2} - k_z^2}}
\right)e^{ik_z z}e^{-i\om t}e^{i n \phi} \right]
\end{array}\label{S2geral2} \ee

with

\bb A_{nm}(k_z,\om) \ug S_{nm}(\om)\delta\left(\om -
(\mathcal{V}k_z + b_m) \right)\label{Anm} \ee

$b_m = 2m\pi \mathcal{V}/\Delta z_0$, and quantity $S_{nm}(\om)$
being an arbitrary frequency spectrum.

\h We should note that, due to Eq.(\ref{Anm}), each term in the
double sum (\ref{S2geral2}), namely in the expression within
square brackets, is a truly nondiffracting wave (beam or pulse),
and their sum (\ref{S2geral2}) is just the most general form
representing an ideal nondiffracting wave defined by
Eq.(\ref{S2def}).

\h We can also notice that (\ref{S2geral2}) is nothing but a
superposition of Bessel beams with a specific space-time coupling
in their spectra: more specifically with linear relationships
between their angular frequency $\om$ and longitudinal wave number
$k_z$.

\h Concerning such a superposition, the Bessel beams with $k_z>0$
($k_z<0$) propagate in the positive (negative) $z$ direction. As
we wish to obtain LWs propagating in the positive $z$ direction,
the presence of ''backward" Bessel beams ($k_z<0$), i.e. of
''backward components", is not desirable. This problem can be
overcome, however, by appropriate choices of the spectrum
(\ref{Anm}), which can totally eliminate those components, or
minimize their contribution, in superposition (\ref{S2geral2}).

\h Another important point refers to the energy of the
LWs\cite{Sez,PIER98,Hillion1,MRH}. It is well known that any ideal
LW, i.e., any field with the spectrum (\ref{Anm}), possesses
infinite energy. However finite-energy LWs can be constructed by
concentrating the spectrum $A_{nm}(k_z,\om)$ in the surrounding of
a straight line of the type $\om = \mathcal{V}k_z + b_m$ instead
of collapsing it exactly over that line. In such a case, the LWs
get a finite energy, but are endowed with finite field depths:
i.e., they maintain their spatial forms for long (but not
infinite) distances.

\h Despite the fact that expression (\ref{S2geral2}), with
$A_{nm}(k_z,\om)$ given by (\ref{Anm}), does represent ideal
nondiffracting waves, it is difficult to use it for obtaining
analytical solutions, especially when having the task of
eliminating the backward components. This difficulty becomes even
worse in the case of finite-energy LWs.

\h As an attempt to bypass these problems, many different
bidirectional and unidirectional decomposition methods have been
proposed in the last years\cite{BSZ,PIER98,MRH}. Those methods
consist essentially in the replacement of the variables $z$ and
$t$ in Eq.(\ref{S2geral2}) with new ones $\zeta = z + v_1t$ and
$\zeta = z + v_2t$, where $v_1$ and $v_2$ are constants possessing
a priori any value in the range $[-\infty,\infty]$. The names
bidirectional and unidirectional decomposition correspond to
$v_1/v_2>0$ and $v_1/v_2<0$, respectively.

\h For instance, in \cite{BSZ} Besieris et.al. introduced the
bidirectional decomposition $\zeta = z-ct$, $\eta = z+ct$ and
obtained interesting ideal and finite energy luminal LWs. They
also worked, in \cite{PIER98}, with the unidirectional
decomposition $\zeta = z-Vt$, $\eta = z - c^2t/V$ (with $V>c$) for
obtaining ideal and finite-energy superluminal LWs. In \cite{MRH}
Zamboni-Rached et.al. introduced the bidirectional decomposition
$\zeta = z-Vt$, $\eta = z+Vt$ (with $V>c$), being thus able to
provide many other ideal and finite-energy superluminal
nondiffracting pulses.

\h Subluminal LWs have been obtained\cite{PIER98,Salo3,pra}, for
instance, through a decomposition of the type $\zeta = z-vt$,
$\eta = z$, with $v<c$.

\h All such decompositions are very efficient in furnishing LWs in
closed forms, but yield solutions that suffer with the problem
that backward travelling wave components enter their spectral
structure. A way out was found in the case of subluminal
waves\cite{pra}, but the problem still persists for the luminal
and superluminal ones. In the latter cases, they succeeded till
now only in minimizing the contribution of the backward Bessel
beams in Eq.(\ref{S2geral2}) by choosing ultra-wideband frequency
spectra. This is not the best approach because, in general, the
solutions found in this way resulted to be far from the
experimental possibilities of the usual laboratories.

\h As it was said before, the backward components can be totally
removed by a proper choice of the spectrum: but none of the
previous decompositions are then able to yield analytical
solutions for the integral (\ref{S2geral2}).

\h We are going to introduce, therefore, a unidirectional
decomposition that allows one to get ideal and finite energy LWs,
with superluminal and luminal peak velocities, without any
occurrence in their spectral structure of backward components.

\subsection{Totally ``forward" LW pulses}

\h Let us start with eqs.(\ref{S2geral1},\ref{S2sepc1}), which
describe a general free-space solution (without evanescent waves)
of the homogeneous wave equation, and consider in
Eq.(\ref{S2sepc1}) a spectrum $A_n(k_z,\om)$ of the type

\bb A_n(k_z,\om) \ug \delta_{n\,0} H(\om)H(k_z)A(k_z,\om)\delta
(k_{\rho}^2 - (\om^2/c^2 - k_z^2)) \label{spec1}\ee

where $\delta_{n\,0}$ is the Kronecker delta function, $H(\cdot)$
the Heaviside function and $\delta(\cdot)$ the Dirac delta
function, quantity $A(k_z,\om)$ being an arbitrary function.
Spectra of the type (\ref{spec1}) restrict the solutions to the
axially symmetric case, with only positive values to the angular
frequencies and longitudinal wave numbers. With this, the
solutions proposed by us get the integral form

\bb \psi(\rho,z,t) \ug \int_{0}^{\infty}\drm\om\,\int_{0}^{\om/c}
\drm k_z \, A(k_z,\om) \, J_0(\rho\sqrt{\om^2/c^2 -
k_z^2})e^{ik_z}e^{-i\om t} \label{superpo1} \ee

i.e., result to be general superpositions of zero-order Bessel
beams propagating in the positive $z$ direction only. Therefore,
any solution obtained from (\ref{superpo1}), be they
nondiffracting or not, are \emph{completely free} from backward
components.

At this point, we introduce the unidirectional decomposition

\bb \left\{\begin{array}{clcr} \zeta \ug z-Vt\\
\\
\eta \ug z-ct\end{array}\right. \label{ze} \ee

with $V>c$.

\h A decomposition of his type was used till now in the context of
paraxial approximation only\cite{paraxial}; but we shall show
below that it can be much more effective, giving important results
in the exact context and in situations that cannot be analyzed in
the paraxial approach.

\h With (\ref{ze}), we can write the integral solution
(\ref{superpo1}) as

\bb \psi(\rho,\zeta,\eta) \ug
(V-c)\int_{0}^{\infty}\drm\sigma\,\int_{-\infty}^{\sigma} \drm
\alpha \, A(\alpha ,\sigma) J_0(\rho\sqrt{\gamma^{-2}\sigma^2
-2(\beta-1)\sigma\alpha}\,\,)\,e^{-i\alpha\eta}e^{i\sigma\zeta}
\label{sup2} \ee

where $\gamma = (\beta^2 -1)^{-1/2}$, $\beta = V/c$ and where

\bb \left\{\begin{array}{clcr} \alpha \ug \dis{\frac{1}{V-c}}\,\,(\om - Vk_z)   \\
\\
\sigma \ug \dis{\frac{1}{V-c}}\,\,(\om - ck_z)
\end{array}\right. \label{as} \ee

are the new spectral parameters.

\h It should be stressed that superposition (\ref{sup2}) is not
restricted to LWs: It is the choice of the spectrum
$A(\alpha,\sigma)$ that will determine the resulting LWs.

\newpage

\subsubsection{Totally ``forward'' ideal superluminal LW pulses}

\textbf{\emph{The X-type waves}}:

\h The most trivial LW solutions are those called X-type
waves\cite{Lu1,Lu2}. They are constructed by frequency
superpositions of Bessel beams with the same phase velocity $V>c$
and \emph{till now} constitute the only known ideal LW pulses free
of backward components. Obviously, it is not necessary to use the
approach developed here to obtain such X-type waves, since they
can be obtained by using directly the integral representation in
the parameters $(k_z,\om)$, i.e., by using Eq.(\ref{superpo1}).
Even so, just as an exercise, let us use the present approach to
construct the ordinary X wave.

\h Consider the spectral function $A(\alpha,\sigma)$ given by

\bb A(\alpha,\sigma) \ug \frac{1}{V-c}\delta(\alpha)e^{-s\sigma}
\label{specx}\ee

\h One can note that the delta function in (\ref{specx}) implies
that $\alpha = 0$ $\rightarrow$ $\om = Vk_z$, which is just the
spectral characteristic of the X-type waves. In this way, the
exponential function ${\rm exp}(-s\sigma)$ represents a frequency
spectrum starting at $\om=0$, with an exponential decay and
frequency bandwidth $\Delta\om=V/s$.

\h Using (\ref{specx}) in (\ref{sup2}), we get

\bb \psi(\rho,\zeta) \ug \frac{1}{\sqrt{(s-i\zeta)^2 +
\gamma^{-2}\rho^2}} \; \equiv X \,\,\, , \label{x} \ee

which is the well known ordinary X wave.

\

\

\textbf{\emph{Totally ``forward" Superluminal Focus Wave Modes}}:

\h Focus wave modes (FWMs)\cite{BSZ,PIER98,MRH} are ideal
nondiffracting pulses possessing spectra with a constraint of the
type $\om = \mathcal{V}k_z + b$ (with $b\neq 0$), which links the
angular frequency and the longitudinal wave number, and are known
for their strong field concentrations.

\h Till now, all the known FWM solutions possess, however,
backward spectral components, a fact that, as we know, forces one
to consider large frequency bandwidths to minimize their
contribution. However we are going to obtain solutions of this
type completely free of backward components, and able to possess
also very narrow frequency bandwidths.

\h Let us choose a spectral function $A(\alpha,\sigma)$ like

\bb A(\alpha,\sigma) \ug \frac{1}{V-c}\delta(\alpha +
\alpha_0)e^{-s\sigma} \label{specfwm}\ee

with $\alpha_0>0$ a constant. This choice confines the spectral
parameters $\om,k_z$ of the Bessel beams to the straight line $\om
= Vk_z - (V-c)\alpha_0$, as it is shown in the figure below

\begin{figure}[!h]
\begin{center}
 \scalebox{3}{\includegraphics{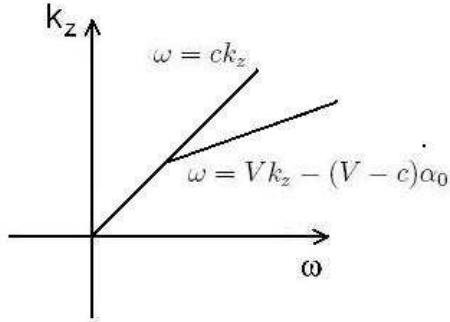}}
\end{center}
\caption{The Dirac delta function in (\ref{specfwm}) confines the
spectral parameters $\om,k_z$ of the Bessel beams to the straight
line $\om = Vk_z - (V-c)\alpha_0$, with $\alpha_0>0$.}
\label{fig1}
\end{figure}

\h Substituting (\ref{specfwm}) in (\ref{sup2}), we have

\bb \psi(\rho,\zeta,\eta) \ug
\int_{0}^{\infty}\drm\sigma\,\int_{-\infty}^{\sigma} \drm \alpha
\, \delta(\alpha + \alpha_0)e^{-s\sigma}
 J_0(\rho\sqrt{\gamma^{-2}\sigma^2
-2(\beta-1)\sigma\alpha})e^{-i\alpha\eta}e^{i\sigma\zeta} \,\,\, ,
\label{intfwm} \ee

which, on using identity $6.616$ in Ref.\cite{grad}, results in

\bb \psi(\rho,\zeta,\eta) \ug X\,e^{i\alpha_0\eta}\,{\rm
exp}\left[\frac{\alpha_0}{\beta+1}\left(s-i\zeta - X^{-1}
\right)\right] \label{fwm}\ee

where $X$ is the ordinary X-wave given by Eq.(\ref{x}).

\h Solution (\ref{fwm}) represents an ideal superluminal LW of the
type FWM, but totally free from backward components.

\h As we already said, the Bessel beams constituting this solution
have their spectral parameters linked by the relation $\om = Vk_z
- (V-c)\alpha_0$; thus, by using (\ref{specfwm}) and (\ref{as}),
it is easy to see that the frequency spectrum of those Bessel
beams starts at $\om_{\rm min} = c\alpha_0$ with an exponential
decay ${\rm exp}(-s\om/V)$, and so possesses the bandwidth
$\Delta\om=V/s$. It is clear that $\om_{\rm min}$ and $\Delta\om$
can assume any values, so that the resulting FWM, eq. (\ref{fwm}),
can range from a quasi-monochromatic to an ultrashort pulse. This
is a great advantage with respect to the old FWM solutions.

\h As an example, we plot two situations related with the LW pulse
given by Eq.(\ref{fwm}).

\begin{figure}[!h]
\begin{center}
 \scalebox{2.3}{\includegraphics{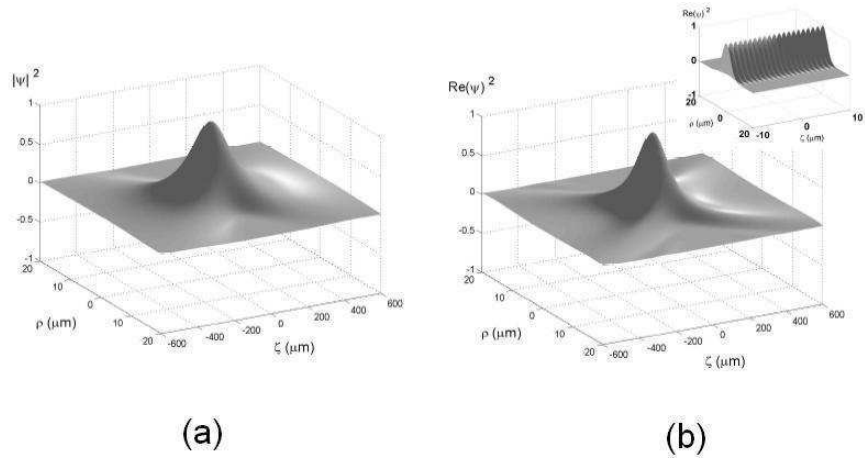}}
\end{center}
\caption{\textbf{(a)} and \textbf{(b)} show, respectively, the
intensity of the complex and real part of a quasi-monochromatic,
totally ``forward", superluminal FWM optical pulse, with
$V=1.5\,c$, $\alpha_0=1.256\times 10^7\,{\rm m}^{-1}$ and $s=
1.194\times 10^{-4}\,{\rm m}$, which correspond to $\om_{\rm
min}=3.77 \times 10^{15}$Hz, and $\Delta\om=3.77 \times
10^{12}$Hz, i.e., to a picosecond pulse with $\lambda_0=0.5\mu$m.}
\label{fig2}
\end{figure}

\h The first, in Fig.(\ref{fig2}), is a quasi-monochromatic
optical FWM pulse, with $V=1.5\,c$, $\alpha_0=1.256\times
10^7\,{\rm m}^{-1}$ and $s= 1.194\times 10^{-4}\,{\rm m}$, which
correspond to $\om_{\rm min}=3.77 \times 10^{15}$Hz, and
$\Delta\om=3.77 \times 10^{12}$Hz, i.e., to a picosecond pulse
with $\lambda_0=0.5\mu$m. Figure (\ref{fig2}a) shows the intensity
of the complex LW field, while Fig.(\ref{fig2}b) shows the
intensity of its real part. Moreover, in Fig.(\ref{fig2}b), in the
right upper corner, it is shown a zoom of this LW, on the $z$ axis
and around the pulse's peak, where the carrier wave of this
quasi-monochromatic pulse shows up.

\begin{figure}[!h]
\begin{center}
 \scalebox{2.3}{\includegraphics{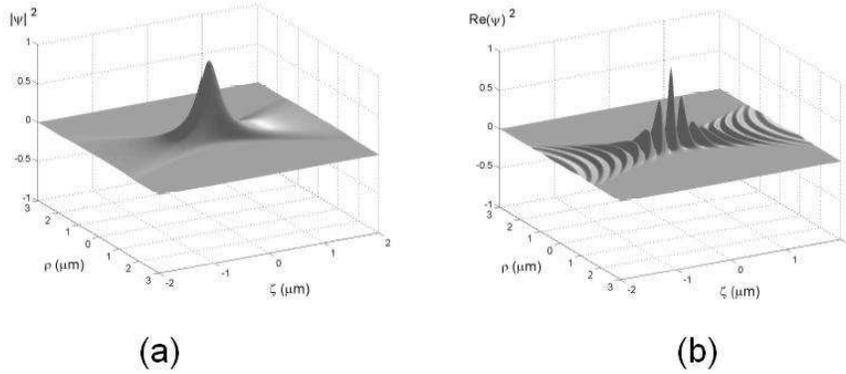}}
\end{center}
\caption{\textbf{(a)} and \textbf{(b)} show, respectively, the
intensity of the complex and real part of an ultrashort, totally
``forward", superluminal FWM optical pulse, with $V=1.5\,c$,
$\alpha_0=1.256\times 10^7\,{\rm m}^{-1}$ and $s=2.3873\times
10^{-7}\,{\rm m}$, which correspond to $\om_{\rm min}=3.77 \times
10^{15}$Hz, and $\Delta\om=1.88\times 10^{15}$Hz, i.e., to a
fentosecond optical pulse.} \label{fig3}
\end{figure}

\h The second example, in Fig.(\ref{fig3}), corresponds to an
ultrashort optical FWM pulse with $V=1.5\,c$,
$\alpha_0=1.256\times 10^7\,{\rm m}^{-1}$ and $s=2.3873\times
10^{-7}\,{\rm m}$, which correspond to $\om_{\rm min}=3.77 \times
10^{15}$Hz, and $\Delta\om=1.88\times 10^{15}$Hz, i.e., to a
fentosecond optical pulse. Figures (\ref{fig3}a,\ref{fig3}b) show
the intensity of the complex and real part of this LW field,
respectively.

\h Now, we apply the present approach to obtain totally ``forward"
finite-energy LW pulses.

\

\subsubsection{Totally ``forward", finite-energy LW pulses}

\h Finite-energy LW pulses are almost nondiffracting, in the sense
that they can retain their spatial forms, resisting to the
diffraction effects, for long (but not infinite) distances.

\h There exist many analytical solutions representing
finite-energy LWs\cite{BSZ,PIER98,MRH}, but, once more, all the
known solutions suffer from the presence of backward components.
We can overcome this limitation.

\h Superluminal finite-energy LW pulses, with peak velocity $V>c$,
can be got by choosing spectral functions in (\ref{superpo1})
which are concentrated in the vicinity of the straight line $\om =
Vk_z + b$ instead of lying on it. Similarly, in the case of
luminal finite-energy LW pulses the spectral functions in
(\ref{superpo1}) have to be concentrated in the vicinity of the
straight line $\om = ck_z + b$ (note that in the luminal case, one
must have $b\geq 0$).

\h Indeed, from Eq.(\ref{as}) it is easy to see that, by our
approach, finite-energy superluminal LWs can be actually obtained
by concentrating the spectral function $A(\alpha,\sigma)$ entering
in (\ref{sup2}), in the vicinity of $\alpha = -\alpha_0$, with
$\alpha_0$ a \emph{positive} constant. And, analogously, the
finite-energy luminal case can be obtained with a spectrum
$A(\alpha,\sigma)$ concentrated in the vicinity of $\sigma =
\sigma_0$, with $\sigma_0 \geq 0$.

\h To see this, let us consider the spectrum

\bb A(\alpha,\sigma) \ug \frac{1}{V-c}H(-\alpha -
\alpha_0)e^{a\alpha}e^{-s\sigma} \label{specfe}\ee

where $\alpha_0>0$, $a>0$ and $s>0$ are constants, and $H(\cdot)$
is the Heaviside function.

\h Due to the presence of the Heaviside function, the spectrum
(\ref{specfe}), when written in terms of the spectral parameters
$\om$ and $k_z$, has its domain in the region shown below:

\begin{figure}[!h]
\begin{center}
 \scalebox{3}{\includegraphics{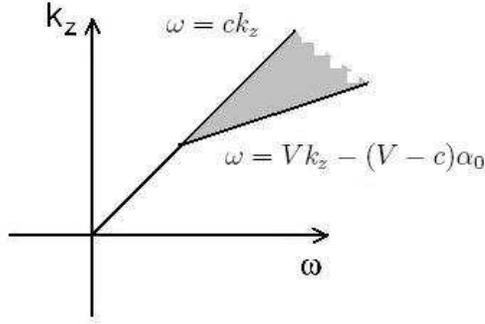}}
\end{center}
\caption{The spectrum (\ref{specfe}), when written in terms of the
spectral parameters $\om$ and $k_z$, has its domain indicated by
the shaded region.} \label{fig4}
\end{figure}

\h We can see that the spectrum $A(\alpha,\sigma)$ given by
Eq.(\ref{specfe}) is more concentrated on the line
$\alpha=\alpha_0$, i.e, around $\om = Vk_z - (V-c)\alpha_0$, or on
$\sigma=0$ (i.e., around $\om = ck_z$), depending on the values of
$a$ and $s$: More specifically, the resulting solution will be a
superluminal finite-energy LW pulse, with peak velocity $V>c$, if
$a>>s$; or a luminal finite-energy LW pulse if $s>>a$.

\h Inserting the spectrum (\ref{specfe}) into (\ref{sup2}), we
have

\bb \psi(\rho,\zeta,\eta) \ug
\int_{0}^{\infty}\drm\sigma\,\int_{-\infty}^{-\alpha_0} \drm
\alpha \, e^{a\alpha}e^{-s\sigma}
 J_0(\rho\sqrt{\gamma^{-2}\sigma^2
-2(\beta-1)\sigma\alpha}\,\,)e^{-i\alpha\eta}e^{i\sigma\zeta}
\,\,\, , \label{intfwm} \ee

and, by using identity $6.616$ in Ref.\cite{grad}, we get

\bb \psi(\rho,\zeta,\eta) \ug X\,\int_{-\infty}^{-\alpha_0} \drm
\alpha e^{a\alpha}\,e^{-i\alpha\eta}\,{\rm
exp}\left[-\frac{\alpha}{\beta+1}\left(s-i\zeta - X^{-1}  \right)
\right]\,\,\, , \ee

which can be directly integrated to furnish

\bb \psi(\rho,\zeta,\eta) \ug \frac{\,\,X\,{\rm
exp}\dis{\left\{-\alpha_0\left[(a-i\eta)-\frac{1}{\beta+1}\left(s-i\zeta
- X^{-1} \right) \right] \right\}}}{\dis{(a-i\eta)
-\frac{1}{\beta+1}\left(s-i\zeta - X^{-1}\right)}}\,\,\, ,
\label{felw}\ee

\h As far as we know, the new solution (\ref{felw}) is the first
one to represent finite-energy LWs completely free of backward
components.

\

\textbf{\emph{Superluminal or luminal peak velocities}}:

\h The finite-energy LW (\ref{felw}) can be superluminal
(peak-velocity $V>c$) or luminal (peak-velocity $c$) depending on
the relative values of the constants $a$ and $s$. To see this in a
rigorous way, in connection with solution (\ref{felw}), we should
calculate how its global maximum of intensity (i.e, its peak),
which is located on $\rho=0$, develops in time. One can obtain the
peak's motion by considering the field intensity of (\ref{felw})
on the $z$ axis, i.e., $|\psi(0,\zeta,\eta)|^2$, at a given time
$t$, and finding out the value of $z$ at which the pulse presents
a global maximum: We shall call $z_{\rm p}(t)$ (the peak's
position) this value of $z$. Obviously the peak velocity will be
$\drm z_{\rm p}(t)/\drm t$.

\h The on-axis field intensity of (\ref{felw}) is

\bb |\psi(0,\zeta,\eta)|^2 \ug
\frac{e^{-2a\alpha_0}}{(as-\eta\zeta)^2 + (s\eta + a\zeta)^2}
\label{oa}\ee

\h For a given time $t$, we can find out the $z$ position of the
peak by setting $\pa |\psi(0,\zeta,\eta)|^2 /\pa z = 0$, that is,

\bb
\frac{e^{-2a\alpha_0}[2(as-\eta\zeta)(\zeta+\eta)-2(s\eta+a\zeta)(a+s)]}{[(as-\eta\zeta)^2
+ (s\eta + a\zeta)^2]^2} \ug 0 \label{c1}\ee

where we have used $\pa/\pa z = \pa / \pa\zeta + \pa/\pa\eta$.

\h From (\ref{c1}), we have that

\bb \eta\zeta^2 + \eta^2\zeta + s^2\eta + a^2\zeta \ug 0
\label{c2} \ee

\h We are interested in the cases where $a>>s$ or $s>>a$.

\h \emph{For the case} $a>>s$, we have for the two last terms in
the l.h.s. of (\ref{c2}): $s^2\eta + a^2\zeta = s^2(z-ct) +
a^2(z-Vt) = (s^2+a^2)z - (s^2c+a^2V)t \approx a^2(z-Vt)$, and so
we can approximate (\ref{c2}) with

\bb \eta\zeta^2 + \eta^2\zeta + a^2\zeta \simeq 0 \label{c3} \ee

which yields three values of $z$. It is not difficult to show that
one of them, $\zeta=0$, i.e., $z=Vt$, furnishes the global maximum
of the intensity (\ref{oa}), and therefore also of (\ref{felw}).
We can also show that the other two roots have real values only
for $t\geq\sqrt{8}\,a/(V-c)$, and in any case furnish values of
the intensity of (\ref{felw}) much smaller than the global
maximum, already found at $z=Vt\equiv z_{\rm p}(t)$. So we can
conclude that for $a>>s$ the peak velocity is $V>c$.

\h\emph{ For the case} $s^2c>>a^2V \rightarrow s>>a$, the two last
terms in the l.h.s. of (\ref{c2}) are: $s^2\eta + a^2\zeta =
s^2(z-ct) + a^2(z-Vt) = (s^2+a^2)z - (s^2c+a^2V)t \approx
s^2(z-ct)$, and so we can approximate (\ref{c2}) with

\bb \eta\zeta^2 + \eta^2\zeta + s^2\eta  \simeq 0 \label{c4} \ee

and we can show that the root $\eta=0$, i.e. $z=ct \equiv z_{\rm
p}(t)$, furnishes the global maximum of the intensity (\ref{oa}),
and therefore also of (\ref{felw}). The other two roots have real
values only for $t\geq\sqrt{8}\,s/(V-c)$, and furnish, once more,
much smaller values of the intensity of (\ref{felw}). In this way,
for $s^2c>>a^2V \rightarrow s>>a$, the peak velocity is $c$.

Now, let us analyze, in details, examples of both cases.

\

\textbf{\emph{Totally ``forward", finite-energy superluminal LW
pulses}}:

\h  As we have seen above, superluminal finite-energy LW pulses
can be obtained from (\ref{felw}) by putting $a>>s$. In this case,
the spectrum $A(\alpha,\sigma)$ is well concentrated around the
line $\alpha=\alpha_0$, and therefore in the plane $(k_z,\om)$
this spectrum starts at $\om_{\rm min} \approx c\alpha_0$ with an
exponential decay, and the bandwidth $\Delta\om \approx V/s$.

\h The field depth of the superluminal LW pulse obtained from
(\ref{felw}), it can be calculated in a simple way. Let us examine
the evolution of its peak intensity by putting $\zeta=0
\rightarrow \eta=(1-\beta^{-1})z$ in eq. (\ref{oa}):

\bb |\psi(0,\zeta=0,\eta)|^2 \ug \frac{e^{-2a\alpha_0}}{a^2s^2 +
s^2\left(1-\beta^{-1}\right)^2z^2} \label{oa2}\ee

where, of course, the coordinate $z$ above is the position of the
peak intensity, i.e., $z=Vt\equiv z_{\rm p}$. It is easy to see
that the pulse presents its maximum intensity at $z=0$ ($t=0$),
and maintains it for $z<<a/(1-\beta^{-1})$. Defining the field
depth $Z$ as the distance over which the pulse's peak intensity
remains at least $25\%$ of its initial value\footnote{We can
expect that, while the pulse peak intensity is maintained, the
same happens for its spatial form.}, we can obtain from
(\ref{oa2}) the depth of field

\bb Z \ug \frac{\sqrt{3}\,a}{1-\beta^{-1}} \ee

which depends on $a$ and $\beta=V/c$: Thus, the pulse can get
large field depths by suitably adjusting the value of parameter
$a$.

\begin{figure}[!h]
\begin{center}
\scalebox{2.5}{\includegraphics{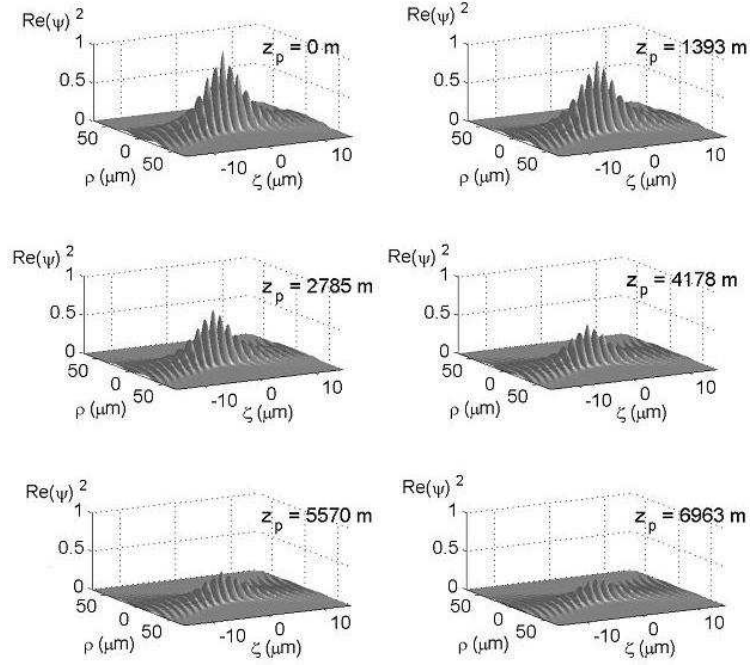}}
\end{center}
\caption{The space-time evolution, from the pulse's peak at
$z_{\rm p}=0$ to $z_{\rm p}=Z$,  of a totally ``forward",
finite-energy, superluminal LW optical pulse represented by
eq.(\ref{felw}), with the following parameter values: $a=20\,$m,
$s=3.99\times 10^{-6}\,$m (note that $a>>s$), $V=1.005\,c$ and
$\alpha_0=1.26\times 10^{7}\,{\rm m}^{-1}$.} \label{fig5}
\end{figure}

\h Figure \ref{fig5} shows the space-time evolution, from the
pulse's peak at $z_{\rm p}=0$ to $z_{\rm p}=Z$,  of a
finite-energy superluminal LW pulse represented by eq.(\ref{felw})
with the following parameter values: $a=20\,$m, $s=3.99\times
10^{-6}\,$m (note that $a>>s$), $V=1.005\,c$ and
$\alpha_0=1.26\times 10^{7}\,{\rm m}^{-1}$. For such a pulse, we
have a frequency spectrum starting at $\om_{\rm min} \approx
3.77\times 10^{15}$Hz (with an exponential decay) and the
bandwidth $\Delta\om \approx 7.54 \times 10^{13}$Hz. From these
values and since $\Delta\om/\om_{\rm min}=0.02$, it is a optical
pulse with $\lambda_0 = 0.5\,\mu$m and time width of $13\,$fs. At
the distance given by the field depth
$Z=\sqrt{3}\,a/(1-\beta^{-1})=6.96\,$km the peak intensity is a
fourth of its initial value. Moreover, it is interesting to note
that, in spite of the intensity decrease, the pulse's spot size
$\Delta\rho_0 = 7.5\,\mu$m remains constant during the
propagation.

\newpage

\textbf{\emph{Totally ``forward", finite-energy luminal LW
pulses}}:

luminal finite energy LW pulses can be obtained from eq.
(\ref{felw}) by making $s>>a$ (more rigorously for $s^2c>>a^2V$).
In this case, the spectrum $A(\alpha,\sigma)$ is well concentrated
around the line $\sigma=0$, and therefore in the plane $(k_z,\om)$
it starts at $\om_{\rm min} \approx c\alpha_0$ with an exponential
decay and the bandwidth $\Delta\om \approx c/a$.

\h The field depth of the luminal LW pulse obtained from
(\ref{felw}) can be calculated in a simple way. Let us examine its
peak intensity evolution by putting $\eta=0 \rightarrow
\zeta=(1-\beta)z$ in eq. (\ref{oa}):

\bb |\psi(0,\zeta=0,\eta)|^2 \ug \frac{e^{-2a\alpha_0}}{a^2s^2 +
a^2\left(1-\beta\right)^2z^2} \label{oa3}\ee

where, of course, the coordinate $z$ above is the position of the
peak intensity, i.e., $z=ct \equiv z_{\rm p}$. It is easy to see
that the pulse has its maximum intensity at $z=0$ ($t=0$), and
maintains it for $z<<s/(\beta-1)$. Defining the field depth $Z$ as
the distance over which the pulse's peak intensity remains at
least $25\%$ of its initial value, we obtain from (\ref{oa3}) the
depth of field

\bb Z \ug \frac{\sqrt{3}\,s}{\beta-1} \ee

which depends on $s$ and $\beta=V/c$.

\h Here, we should note that the bigger the value of $s$, the
smaller the transverse field concentration of the luminal pulse.
This occurs because for big values of $s$ the spectrum becomes
strongly concentrated around $\om=ck_z$ and, as one knows, in this
case, the solution tends to become a plane wave pulse\footnote{A
possible solution for this limitation it would be the use of
spectra concentrated around the line $\sigma=\sigma_0>0$.}.

\h Let us consider, for instance, a finite-energy luminal LW pulse
represented by eq.(\ref{felw}) with $a=1.59\times 10^{-6}$m,
$s=1\times 10^{4}$m (note that $s>>a$), $V=1.5\,c$,
$\alpha_0=1.26\times 10^{7}\,{\rm m}^{-1}$. For such a pulse,
which has its peak travelling with the light velocity $c$, the
frequency spectrum starts at $\om_{\rm min} \approx 3.77\times
10^{15}$Hz with a bandwidth $\Delta\om \approx 1.88\times
10^{14}$Hz. Thus, it is a optical pulse with time width of
$5.3\,$fs.

\begin{figure}[!h]
\begin{center}
\scalebox{2.5}{\includegraphics{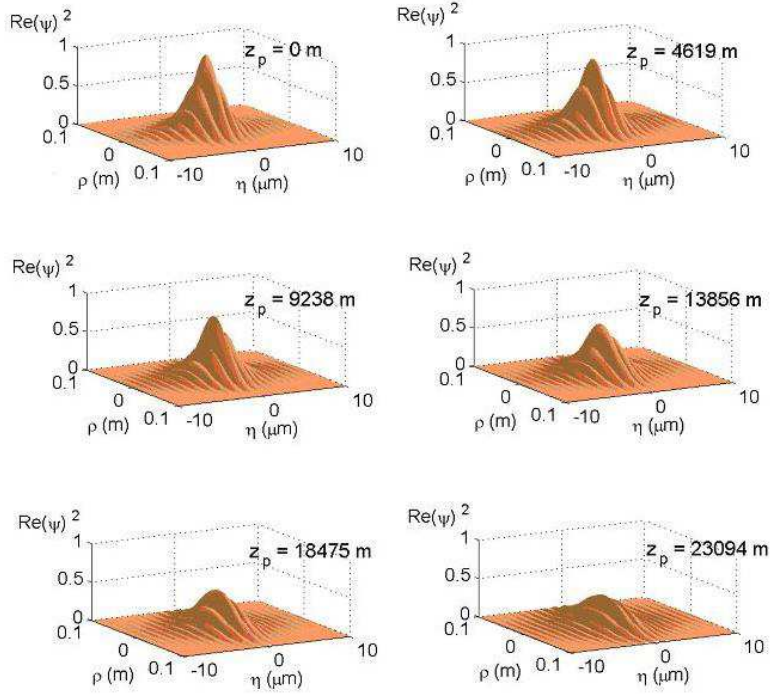}}
\end{center}
\caption{The space-time evolution, from the pulse's peak at
$z_{\rm p}=0$ to $z_{\rm p}=Z$,  of a totally ``forward",
finite-energy, luminal LW optical pulse represented by
eq.(\ref{felw}), with $a=1.59\times 10^{-6}$m, $s=1\times 10^{4}$m
(note that $s>>a$), $V=1.5\,c$, $\alpha_0=1.26\times 10^{7}\,{\rm
m}^{-1}$.} \label{fig6}
\end{figure}

\h The space-time evolution of this pulse, from $z_{\rm p}=0$ to
$z_{\rm p}=Z$, is shown in Figure (\ref{fig6}). At the distance
given by the field depth $Z=\sqrt{3}\,s/(\beta-1)=23.1\,$km the
peak intensity is a fourth of its initial value.

\h We can see from these two examples, and it can also be shown in
a rigorous way, that the superluminal LW pulses obtained from
solution (\ref{felw}) are superior than the luminal ones obtained
from the same solution, in the sense that the former can possess
large field depths and, at same time, present strong transverse
field concentrations. To obtain more interesting and efficient
luminal LW pulses we should use spectra concentrated around the
line $\sigma=\sigma_0>0$.

\

\

\subsubsection{Nonaxially symmetric, totally ``forward", LW pulses}

\h So far, we have applied the present method to axially symmetric
solutions only.

\h A simple way for obtaining the nonaxially symmetric versions of
the previous LW pulses is through the following superposition

\bb \psi(\rho,\phi,\zeta,\eta) \ug
(V-c)\int_{0}^{\infty}\drm\sigma\,\int_{-\infty}^{\sigma} \drm
\alpha \, A'(\alpha ,\sigma) J_{\nu}(\rho\sqrt{\gamma^{-2}\sigma^2
-2(\beta-1)\sigma\alpha}\,\,)\,e^{i\nu\phi}\,e^{-i\alpha\eta}e^{i\sigma\zeta}
\label{sup3} \ee

where $\phi$ is the azimuth angle, $\nu$ is an integer and

\bb A'(\alpha ,\sigma) \ug \left(\frac{\sigma}{\sigma -
2\alpha/(\beta+1)} \right)^{\nu/2}\,A(\alpha ,\sigma)
\label{alinha} \ee

with $A(\alpha ,\sigma)$ being the respective spectra of the
axially symmetric LW solutions of the previous subsections, which
can be recover from Eq.(\ref{sup3}) by making $\nu=0$. Thus, the
fundamental superposition (\ref{sup3}) is more general than
(\ref{sup2}), in the sense that the former can yield both, axially
symmetric and nonaxially symmetric LWs totally free of backward
components.

\h \textbf{\emph{Totally ``forward", nonaxially LW of the type
FWM}} can be reached from Eq.(\ref{sup3}) by using (\ref{specfwm})
into (\ref{alinha}). After integrating directly over $\alpha$, the
integration over $\sigma$ can be made by using the identity
$3.12.5.6$ in Ref.\cite{prud}, furnishing:

\bb \psi(\rho,\phi,\zeta,\eta) \ug e^{i\nu\phi}
\left(\frac{\gamma^{-1}\rho}{s-i\zeta + X^{-1}}\right)^{\nu}
X\,e^{i\alpha_0\eta}\,{\rm
exp}\left[\frac{\alpha_0}{\beta+1}\left(s-i\zeta - X^{-1}
\right)\right] \label{fwm2}\ee

\

\h \textbf{\emph{Totally ``forward", finite-energy, nonaxially
LW}} is obtained from Eq.(\ref{sup3}) by using (\ref{specfe}) into
(\ref{alinha}). By integrating first over $\sigma$ (with the
identity $3.12.5.6$  in Ref.\cite{prud}), the integration over
$\alpha$ can be made directly, and Eq.(\ref{sup3}) yields

\bb \psi(\rho,\phi,\zeta,\eta) \ug e^{i\nu\phi}
\left(\dis{\frac{\gamma^{-1}\rho}{s-i\zeta +
X^{-1}}}\right)^{\nu}\,\frac{\,X\,{\rm
exp}\dis{\left\{-\alpha_0\left[(a-i\eta)-\frac{1}{\beta+1}\left(s-i\zeta
- X^{-1} \right) \right] \right\}}}{\dis{(a-i\eta)
-\frac{1}{\beta+1}\left(s-i\zeta - X^{-1}\right)}}\,\,\, ,
\label{felw2}\ee

\

\h The solution above can possesses superluminal ($a>>s$) or
luminal ($s>>a$) peak-velocity.

\

\

\subsubsection{A functional expression for totally ``forward" LW pulses}

\h In the literature concerning the LWs\cite{BandS} some
interesting approaches appear, capable of yielding functional
expressions which describe LWs in closed form. Although
interesting, even the LWs obtained from those approaches also
possess backward components in their spectral structure.

\h Now, however, we are able to obtain a kind of functional
expression capable of furnishing totally ``forward" LW pulses.

\h Let us consider, in eq.(\ref{sup3}), spectral functions
$A'(\alpha,\sigma)$ of the type

\bb A'(\alpha,\sigma) \ug
(V-c)\,H(-\alpha)\left(\frac{\sigma}{\sigma - 2\alpha/(\beta+1)}
\right)^{\nu/2}\Lambda'(\alpha)\,e^{-s\sigma} \label{gs} \ee

where $H(.)$ is, as before, the Heaviside function, and
$\Lambda'(\alpha)$ a general function of $\alpha$.

\h Using (\ref{gs}) in (\ref{sup3}), performing the integration
over $\sigma$ and making the variable change $\alpha = -u$, we get

\bb \psi(\rho,\phi,\zeta,\eta) \ug e^{i\nu\phi}
\left(\frac{\gamma^{-1}\rho}{s-i\zeta + X^{-1}}\right)^{\nu}
X\,\int_{0}^{\infty} \drm u \, \Lambda(u)\,e^{-u \, S}\label{sup4}
\ee

quantity $X$ being the ordinary X-wave (\ref{x}) and $S$ being
given by

\bb S \ug -i\eta - \frac{1}{\beta +1}(s-i\zeta - X^{-1}) \ee

\h The integral in (\ref{sup4}) is nothing but the Laplace
transform of $\Lambda(u)$.

In this way, we have that

\bb \psi(\rho,\phi,\zeta,\eta) \ug e^{i\nu\phi}
\left(\frac{\gamma^{-1}\rho}{s-i\zeta +
X^{-1}}\right)^{\nu}X\,F\left(-i\eta - \frac{1}{\beta +1}(s-i\zeta
- X^{-1})\right)\,\, , \label{glw} \ee

with $F(.)$ an arbitrary function, is an exact solution to the
wave equation that can yield ideal, and also finite-energy LW
pulses, with superluminal or luminal peak velocities. Besides
this, if the chosen function $F(S)$ in Eq.(\ref{glw}) is regular
and free of singularities at all space-time points
$(\rho,\phi,z,t)$, we can show that the LW solutions obtained from
Eq.(\ref{glw}) will be totally free of backward components.

\

\section{Conclusions}

\h In conclusion, by using a unidirectional decomposition we were
able to get totally ``forward", ideal and finite-energy LW pulses.
These new solutions are superior than the known LWs already
existing in the literature, since the old solutions suffer from
the undesirable presence of backward components in their spectra.

\h By overcoming the problem of these noncausal components, the
new LWs here obtained are not obliged to have physical sense only
in the cases of ultra-wideband frequency spectra; actually, our
new LWs can also be quasi monochromatic pulses, and in such a way
get closer to a practicable experimental realization.

\

\section{Acknowledgements}

The author is very grateful to Erasmo Recami, I.M.Besieris, Hugo
E. Hern\'andez-Figueroa and Claudio Conti  for continuous
discussions and kind collaboration. Thanks are also due to Jane M.
Madureira and Suzy Z. Rached.

\h This work was supported by CNPq and FAPESP (Brazil).

\h E-mail address: mzamboni@ufabc.edu.br

\

\

\end{document}